\documentclass[%
prx,
 amsmath,amssymb,
 reprint,%
]{revtex4-2}
\usepackage{graphicx}
\usepackage{dcolumn}
\usepackage{bm}
\usepackage{color}
\usepackage{float}

\usepackage[utf8]{inputenc}
\usepackage[T1]{fontenc}
\usepackage{mathptmx}
\usepackage{subfig}
\usepackage[inline]{enumitem}
\usepackage{caption}
\newcommand{\bogus}[1]{{}}

\begin{document}

\title{Staged cooling of a fusion-grade plasma in a tokamak thermal quench}

\author{Jun Li}%
\affiliation{Theoretical Division, Los Alamos National Laboratory, Los Alamos, New Mexico 87545, USA}
\affiliation{School of Nuclear Science and Technology, University of Science and Technology of China, Hefei, Anhui 230026, China~\footnote{Present address}}
\author{Yanzeng Zhang}%
\affiliation{Theoretical Division, Los Alamos National Laboratory, Los Alamos, New Mexico 87545, USA}
\author{Xian-Zhu Tang}%
\affiliation{Theoretical Division, Los Alamos National Laboratory, Los Alamos, New Mexico 87545, USA}


\begin{abstract}
In tokamak disruptions where the magnetic connection length becomes
comparable to or even shorter than the plasma mean-free-path, parallel
transport can dominate the energy loss and the thermal quench of the
core plasma goes through four phases (stages) that have distinct
temperature ranges and durations.  The main temperature drop occurs
while the core plasma remains nearly collisionless, with the parallel
electron temperature $T_{e\parallel}$ dropping in time $t$ as
$T_{e\parallel}\propto t^{-2}$ and a cooling time that scales with the
ion sound wave transit time over the length of the open magnetic field
line. These surprising physics scalings are the result of effective
suppression of parallel electron thermal conduction in an otherwise
bounded collisionless plasma, which is fundamentally different from
what are known to date on electron thermal conduction along the
magnetic field in a nearly collisionless plasma.
\end{abstract}

\maketitle

\section{Introduction}

The energy confinement time of a fusion-grade plasma in an ITER-like
tokamak reactor is around 1 second (s).~\cite{ITER-transport-nf-2007}
This is set by plasma transport across magnetic field lines, or more
accurately, nested magnetic surfaces.~\cite{boozer-rmp-2004} When a
major disruption occurs, such a plasma is expected to lose most of its
thermal energy over a period of around 1 millisecond (ms) or so, as
noted in the original ITER design document.~\cite{Hender-etal-nf-2007}
A dedicated experimental campaign on JET~\cite{riccardo2005timescale},
after the initial ITER design was
completed~\cite{iter-basis-chapter3-nf-1999}, revealed that the
thermal quench (TQ) duration could vary greatly, from 0.05 to 3.0~ms.
The significance of a tokamak thermal quench is that it marks the
point of no return in a tokamak disruption.  It not only brings a
thermal load management issue at the divertor plates and first wall,
but also determines the runaway seeding for the subsequent current
quench as well as the parallel electric field that could drive
avalanche growth of runaway electrons.  Since the faster a thermal
quench is, the more problematic it becomes for effective mitigation,
there are practical interests and urgencies in understanding the
fundamentals of the transport physics that are responsible for the
rapid thermal quench, particularly the large variation in its
duration.

The 3-4 order of magnitude faster plasma energy loss rate in a tokamak
disruption~\cite{riccardo2005timescale,shimada2007chapter,nedospasov2008thermal}
is intuitively understood as the result of parallel transport along
open magnetic field lines dominating over the perpendicular transport,
which is across the magnetic field lines. For that to materialize, the
magnetic field lines must connect the hot fusion plasma of 10-15~KeV
temperature directly to a cold and dense plasma, which would also
serve as a radiative energy sink. This is known to occur in at least
two scenarios. The first and more common one of the naturally
occurring disruptions has the globally stochastic magnetic field lines
connecting the hot core plasma directly onto the divertor surface
and/or the first wall, as the result of large-scale
magnetohydrodynamic instabilities destroying nested magnetic
surfaces.~\cite{bondeson1991mhd,riccardo2002disruption,nardon2016progress,sweeney2018relationship}
The other is high-Z impurity injection in the form of deliberately
injected solid pellets or accidentally falling tungsten
debris.~\cite{federici2003key,combs2010alternative,commaux2016first,paz2020runaway}
In both situations, a nearly collisionless plasma is made to intercept
a radiative cooling mass, being that an ablated pellet or a
vapor-shielded wall.

The simplest problem setup to decipher the parallel cooling physics is
to unwind the open field lines into a slab and have a hot plasma of
temperature $T_0$ bounded by thermobath boundary at the two ends that
recycles plasma particles but clamps the temperature of the recycled
plasma particles to $T_w\ll T_0.$ The length of the slab or field
line, $x\in[-L_B,L_B],$ is twice the magnetic connection length $L_B.$
The other characteristic length of the problem is the parallel
mean-free-path $\lambda_{mfp},$ which is tens of kilometers for a
plasma of $T_0=$10-15~KeV and $n_e=10^{19-20}$m$^{-3}.$ The ratio of
the two is the Knudsen number $K_n\equiv \lambda_{mfp}/L_B,$ and it
sets the different cooling regimes (stages) that a thermal quench can
go through. For an initial plasma of $K_n\ge 1,$ the TQ would start
with the (nearly) collisionless phase (regime) and eventually
transition to the collisional phase after the plasma cools down
sufficiently as $K_n\propto
T_e^2/n_e.$~\cite{gray1980measurement,Bell-pof-1985,shvarts1981self,landi2001temperature}
The simplicity of this prototypical problem setup reinforces
the general importance and broad applicability of the underlying
plasma cooling physics in fusion as well as non-fusion applications,
which will be shown in this paper to present a number of surprises.

The prevailing view on plasma thermal quench is the dominant role of
electron {\em parallel thermal conduction}. This is described in the
collisional limit by Braginskii closure for parallel electron
conduction flux~\cite{braginskii}
\begin{align}
  q_{e\parallel} \equiv \int m_e \mathbf{\tilde{v}}^2\tilde{v}_\parallel f_e
  d^3\mathbf{v}=- 3.16\frac{n_eT_e\tau_{e}}{m_e}\frac{\partial
    T_e}{\partial x}\sim
  n_ev_{th,e}T_e\frac{\lambda_{mfp}}{L_B},\label{eq-qen-Braginskii}
\end{align}
with $\tau_e$ the electron collision time, $m_e$ the electron mass,
$v_{th,e}=\sqrt{T_e/m_e}$ the electron thermal speed, and
$\mathbf{\tilde{v}}=\mathbf{v}-\mathbf{V}_e$ the peculiar velocity.
As the collisionality reduces, $q_{e\parallel}$ scales up linearly
with $K_n,~ q_{e\parallel}\propto K_n.$ This would start to break down
when $K_n$ gets above $10^{-2}$ or so, and when $K_n\sim 1$ or $K_n\gg 1, q_{e\parallel}$
would {\em retain the $n_e v_{th,e} T_e$ scaling} but the
$K_n$ term is replaced by a saturated numerical factor $\alpha \sim
0.1,$~\cite{Bell-pof-1985}
\begin{align}
q_{e\parallel} = \alpha n_e v_{th,e} T_e. \label{eq:qe-free-streaming}
\end{align}
With the so-called flux-limiting form of
Eq.~(\ref{eq:qe-free-streaming}) in the collisionless or
free-streaming regime~\cite{atzeni_book_2004},
one finds the solution of the heat conduction equation in the bounded domain of $x\in [-L_B, L_B]$ showing a thermal quench duration
\begin{align}
\tau_{TQ} \sim L_B/v_{th,e}. \label{eq:TQ-scaling-free-streaming}
\end{align}
This suggests a very fast thermal quench indeed, on the order of the
thermal electron   ma over the open magnetic
field line, for a tokamak plasma having $K_n\sim 1$ or $K_n\gg 1$ at
the onset of a disruption.

\begin{figure}[h!]
\centering
\includegraphics[width=0.45\textwidth]{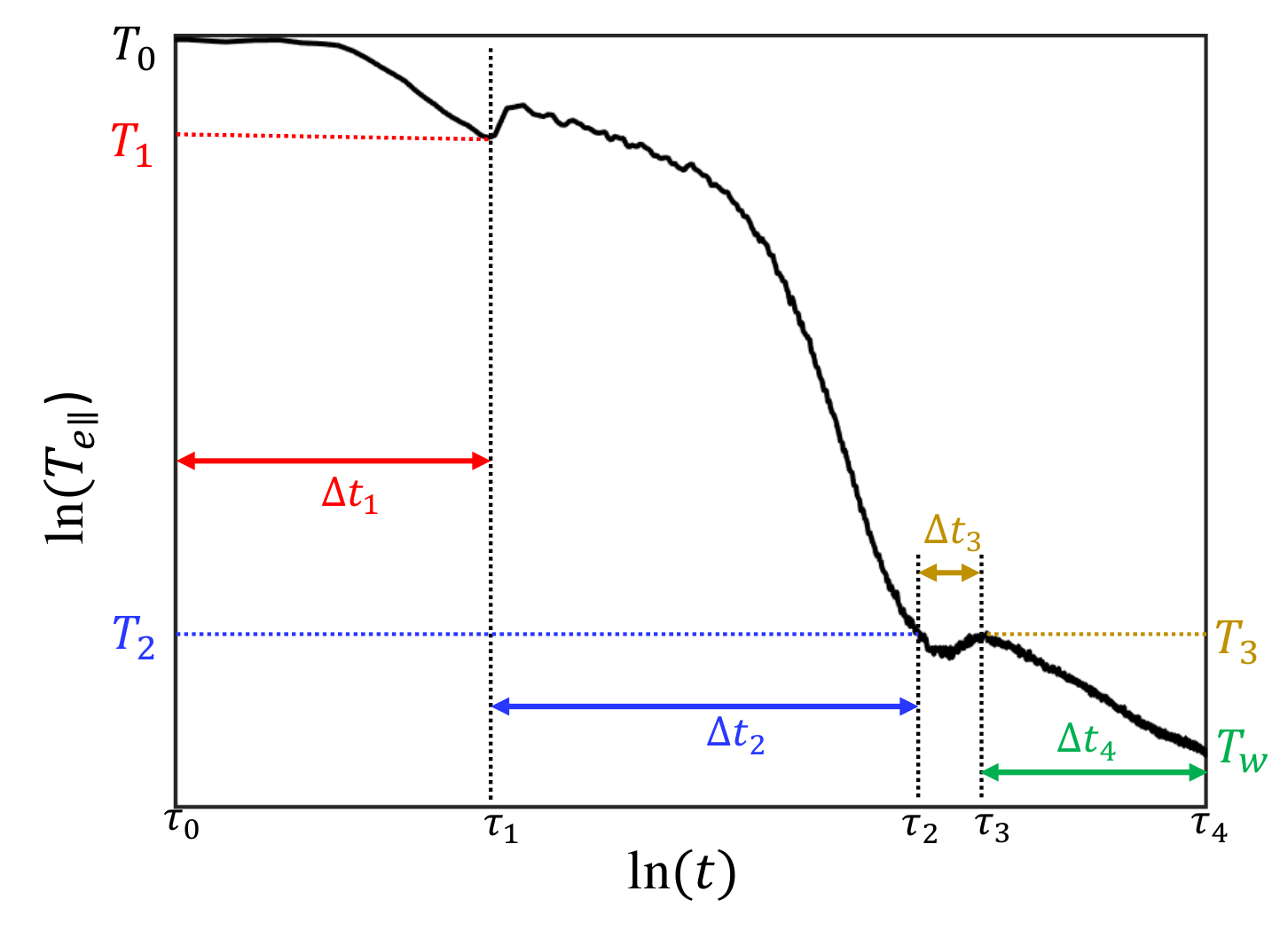}
\caption{Schematic view of core $T_{e\parallel}$ evolution in log-log scale from VPIC simulations corresponding to Fig.~\ref{fig:Te_Ti_para_perp}. }
\label{fig:Te_para-phases}
\end{figure}

Here we show that the core plasma cooling in a bounded
plasma like that of a tokamak disruption follows {\em qualitatively and
  quantitatively different} behaviors. Specifically the parallel
electron temperature $T_{e\parallel}(x=0,t)$ goes through four
distinct stages with durations $\Delta t_{1,2,3,4}$ segmented by
transition temperatures of $T_{1,2,3},$ as illustrated in
Fig.~\ref{fig:Te_para-phases}. There is a precursor stage of very
short duration at the electron transit time $\Delta t_1 \sim
\tau_{tr}^e\equiv L_B/v_{th0}$ with $v_{th0}\equiv \sqrt{T_0/m_e}$ the
initial electron thermal speed, which follows the same scaling as in
Eq.~(\ref{eq:TQ-scaling-free-streaming}). Interestingly it produces
rather limited cooling of the core plasma, $T_1 \ge 0.6 T_0 \gg T_w.$
The core plasma cooling is primarily accomplished in the next
collisionless stage from $T_1 (\sim T_0)$ to $T_2 (\ll T_0),$ with
$\Delta t_2 \gg \Delta t_1.$ There is a transition stage $\Delta t_3
(\sim \Delta t_2)$ that connects the collisionless cooling phase to
the eventual collisional cooling phase, in which $\Delta t_4$ is far
longer than the main collisionless cooling stage of $\Delta t_2.$ The
{\em most surprising and impactful finding} is that the main
collisionless cooling phase from $T_1\sim T_0$ to $T_2 \ll T_0,$ is
dominated by convective energy transport as opposed to the commonly
accepted much faster electron parallel thermal conduction, e.g.,
Eq.~(\ref{eq:qe-free-streaming}).  This yields a thermal quench time
$\Delta t_2$ that {\em scales with the ion transit time} $\Delta t_2
\propto \tau_{tr}^i \equiv L_B/c_s$ with $c_s\equiv
\sqrt{3(1+Z)T_0/m_i}$ the ion sound speed ($Z$ is the ion charge). The
main thermal quench time of $\Delta t_2$ is thus {\em qualitatively
  different} from that of $\tau_{TQ}$ in
Eq.~(\ref{eq:TQ-scaling-free-streaming}), which constitutes the
prevailing textbook description of thermal loss by electron
free-streaming along magnetic field lines.

The rest of the paper is organized as follows. In
Section~\ref{sec:methods}, we explain the setup of the
first-principles kinetic simulations.  The simulation results are
analyzed in Section~\ref{sec:results}, where the physics underlying
the four distinct phases of plasma cooling is elucidated with
theory and modeling. The paper is concluded with a summary of results
and additional discussions for current and planned tokamak experiments
in Section~\ref{sec:conclusion}. There is also an
Appendix~\ref{appendix:collisional-cooling} that gives the derivation
of the collisional cooling scaling.

\section{Methods\label{sec:methods}}

\textbf{Simulation setup:} We deployed first-principles kinetic
simulations using the VPIC code~\cite{VPIC} to investigate the cooling
dynamics of an open field line plasma.  The simulation domain is 1-D
with domain size of $1400\lambda_{De}$, where $\lambda_{De}$ is the
Debye length for the initial plasma conditions with uniform electron
and ion density $n_0=1\times 10^{19}cm^{-3}$ and temperature
$T_0=10keV$. We use cell size $\Delta x=0.1\lambda_{De}$ with 1000
electrons and 1000 hydrogen ions per cell. The time step $\Delta
t=1.357\times 10^{-2} \omega_{pe}^{-1}$ is chosen according to the
Courant condition.  To elucidate the underlying physics, we perform
two contrasting simulations with (1) thermobath boundary condition
that models the cooling effect of a radiative cooling mass, and (2) an
absorbing boundary.  The thermobath boundary recycles particles by
re-injecting electron-ion pairs with a clamped temperature
$T_w=0.01T_0$, while the absorbing boundary will absorb all the
particles hitting the boundary.

\section{Four distinct phases of plasma cooling\label{sec:results}}

\begin{figure}[h!]
\centering
\includegraphics[width=0.45\textwidth]{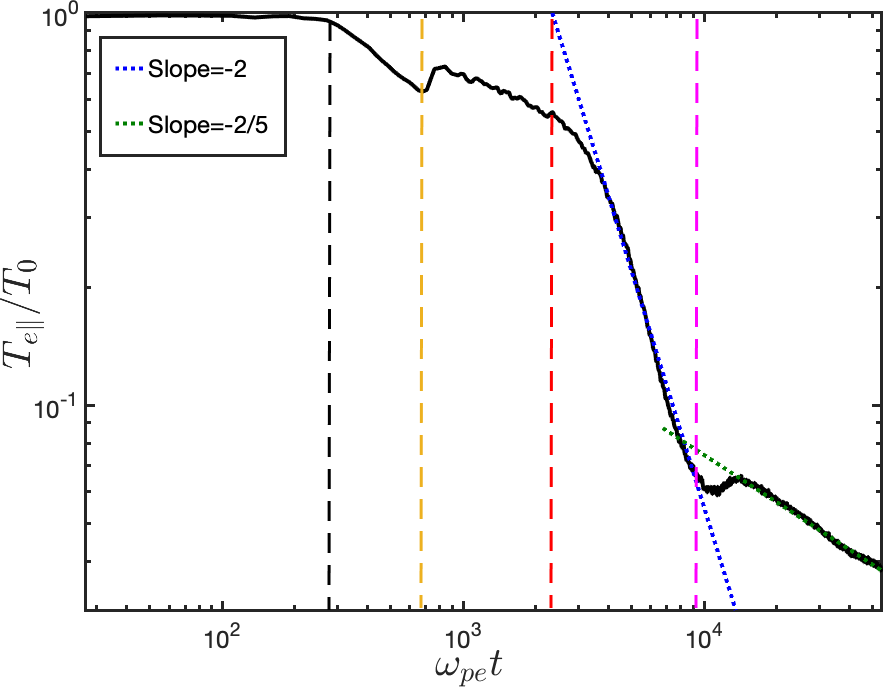}
\caption{$T_{e\parallel}(x=0,t)$ of a slab ($x\in
  [-L_B, L_B]$) is shown for a VPIC simulation with
  $L_B=1400\lambda_{De}$ ($\lambda_{De}$ the electron Debye length),
  an initially uniform plasma of $m_e/m_i=1/100, Z=1$ and $K_{n,0}=37$.
  Thermobath boundary on two ends has $T_w=0.01 T_0.$ The vertical dashed
  lines mark the moments when the four fronts reach the center, while
  the dotted lines with slopes of $-2$ (blue) and $-2/5$ (green) are
  marked for the cooling flow phase ($\Delta t_2$) and collisional
  phase ($\Delta t_4$).}
\label{fig:Te_Ti_para_perp}
\end{figure}

{\em Precursor phase ($T_{e\parallel}$ from $T_0$ to $T_1$ and $\Delta
  t_1 \sim \tau_{tr}^e$):} At the onset of a plasma thermal quench
when the hot plasma on the open field lines suddenly intercepts a cold
boundary, $T_{e\parallel}(x=0,t)$ barely changes over a period that is
approximately one third of $\tau_{tr}^e $. This is simply the time for
the so-called precooling front (PF), which is driven by suprathermal
electron loss and propagates from the boundary toward the core plasma
at a speed $U_{PF}=2.4 v_{th0}$, to arrive at the plasma center
($x=0$).~\cite{Zhangfronts} This instance is marked by the black
dashed vertical line in Fig.~\ref{fig:Te_Ti_para_perp}.  The
subsequent cooling of $T_{e\parallel}(x=0)$ is indeed driven by the
electron conduction flux that follows a similar scaling as the flux
limiting form in Eq.~(\ref{eq:qe-free-streaming}). Notice that for a
nearly collisionless plasma, the conduction flux corresponding to
$T_{e\parallel}$ cooling is given by $q_{en} \equiv \int m_e
\tilde{v}_\parallel^3 f_e d^3\mathbf{v}$. Consequently, the duration
of this phase does follow the free-streaming scaling previously given
in Eq.~(\ref{eq:TQ-scaling-free-streaming}).  Remarkably the precursor
phase ends abruptly when the so-called precooling trailing front (PTF)
reaches the center. This is a second electron front that propagates
from the boundary toward the core plasma, with a speed
$U_{PTF}=\sqrt{2e\Delta\Phi_{RF}/m_e}.$~\cite{Zhangfronts} Here the
reflecting potential $\Delta\Phi_{RF}$ is the result of the ambipolar
electric field in the ion recession layer where a plasma rarefaction
wave is formed.  Since the ambipolar potential scales with
$T_{e\parallel},$ we would normally have $v_{th,e} < U_{PTF} <
U_{PF},$ and thus $\Delta t_1 \sim \tau_{tr}^e.$ The short duration of
the precursor phase produces very limited cooling and typically $T_1
\ge 0.6 T_0.$

{\em Cooling flow phase ($T_{e\parallel}$ from $T_1$ to $T_2$ with $T_w <
  T_2 \ll T_0$ and $\Delta t_2 \sim \tau_{tr}^i$):} Once the
precooling trailing fronts reach the plasma center from both ends,
$q_{en}$ undergoes a qualitative transition from scaling with
$v_{th,e}$ (i.e. $q_{en} \propto n_e v_{th,e} T_{e\parallel}$)
to scaling with $V_{i\parallel}$ (i.e. $q_{en}\propto n_e V_{i\parallel}
T_{e\parallel}$). Here $V_{i\parallel}$ is the parallel ion convective
flow and ambipolar transport implies the parallel electron flow
$V_{e\parallel} \approx V_{i\parallel}.$ Previously we have shown that
in a semi-infinite plasma bounded at one side only by a thermobath
boundary, $q_{en}$ always has the free-streaming flux-limiting
form $n_e v_{th,e} T_{e\parallel}$ due to the cold electrons being pulled into
the hot plasma by the ambipolar electric field.~\cite{Zhangfronts}  The qualitative change of $q_{en}$ in a bounded plasma is driven by how
$T_{e\parallel}$ is being cooled in a long mean-free-path plasma where
electron trapping is provided by the reflecting (ambipolar) potential
along the magnetic field line on both ends.

The core cooling physics is most straightforwardly understood in a
plasma with perfectly absorbing boundaries. Since the trapped-passing
boundary is given by $v_c\equiv\sqrt{e\Delta\Phi_{RF}/m_e},$ and the
truncated electron distribution at $x=0$ has the form
\begin{align}
f_0 = \frac{n_0}{(2\pi)^{3/2}
  v_{th0}^3} e^{-\left(v_\parallel^2+v_\perp^2\right)/2v_{th0}^2}
\Theta\left(1-\frac{v_\parallel}{v_c}\right)
\Theta\left(1+\frac{v_\parallel}{v_c}\right), \nonumber
\end{align}
with $\Theta(x)$ the Heaviside function satisfying $\Theta(x<0)=0$
and $\Theta(x>0)=1.$ One can see that the parallel electron
temperature $ T_{e\parallel}(v_c,T_0) =\int m_e \tilde{v}_\parallel^2 f_e
  d^3\mathbf{v}/\int f_e
  d^3\mathbf{v}= \left(v_c^2/3
v_{th0}^2\right) T_0 $ of the electrostatically trapped long
mean-free-path electrons cools drastically with a decreasing
$v_c/v_{th0} \ll 1.$ This correlates with a reducing reflecting
potential $\Delta\Phi_{RF}.$ Away from the symmetry point at $x=0,$ a
finite parallel electron conduction flux arises from the asymmetric
truncations ($V_L$ and $V_R$) on the positive and negative
 sides of electron distribution in $v_\parallel$
\begin{align}
f_e = \frac{n_0}{(2\pi)^{3/2}
  v_{th0}^3} e^{-\left(v_\parallel^2+v_\perp^2\right)/2v_{th0}^2}
\Theta\left(1-\frac{v_\parallel}{V_R}\right)
\Theta\left(1+\frac{v_\parallel}{V_L}\right). \nonumber
\end{align}
In the limit of significant plasma cooling, one can write $
V_{e\parallel} \approx \left(-V_L+V_R\right)/2 $ and find a convective
energy flux scaling for electron thermal conduction 
\begin{align}
  q_{en} \approx \frac{6T_{e\parallel}}{5T_0} n_e V_{e\parallel}
  T_{e\parallel}. \label{eq:qn-convective-scaling}
\end{align}
In a cooling plasma of $T_{e\parallel}\ll T_0,$ one recovers the
remarkable result of $q_{en} \ll n_e V_{e\parallel}
T_{e\parallel}$ observed in VPIC simulations.

\begin{figure*}[tbh]
\centering
\includegraphics[width=0.4\textwidth]{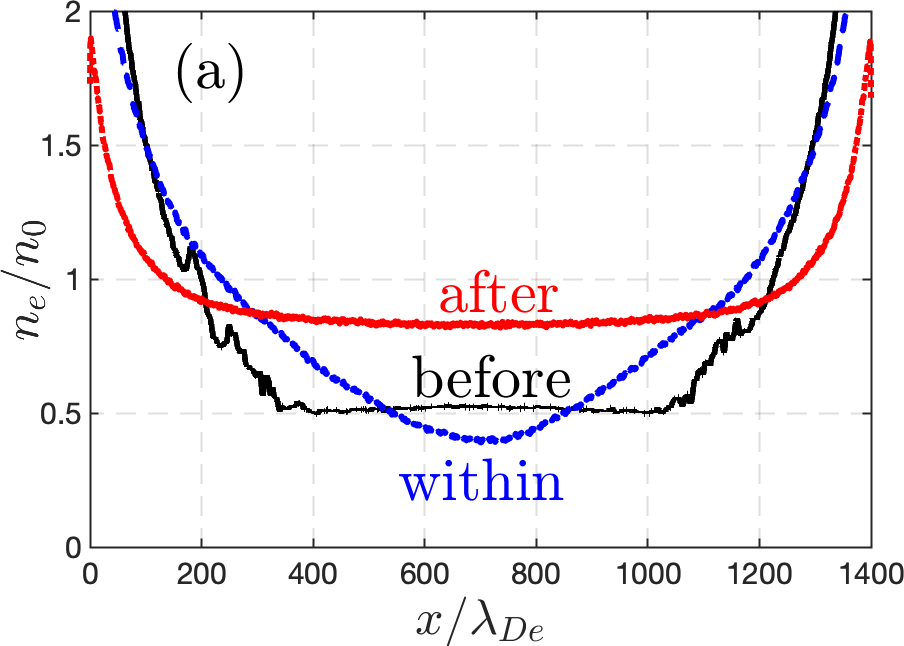}
\includegraphics[width=0.4\textwidth]{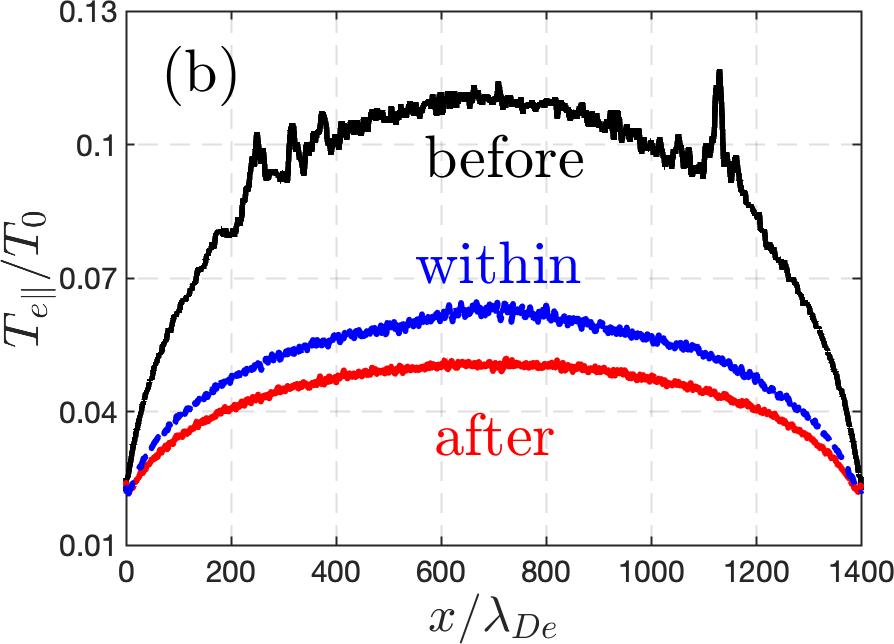}\\
\includegraphics[width=0.4\textwidth]{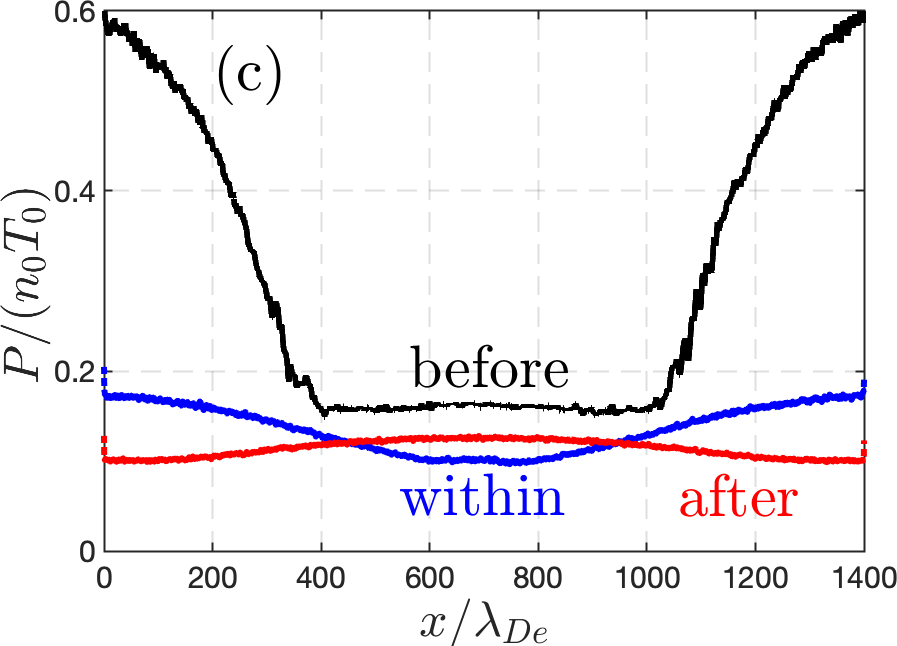}
\includegraphics[width=0.4\textwidth]{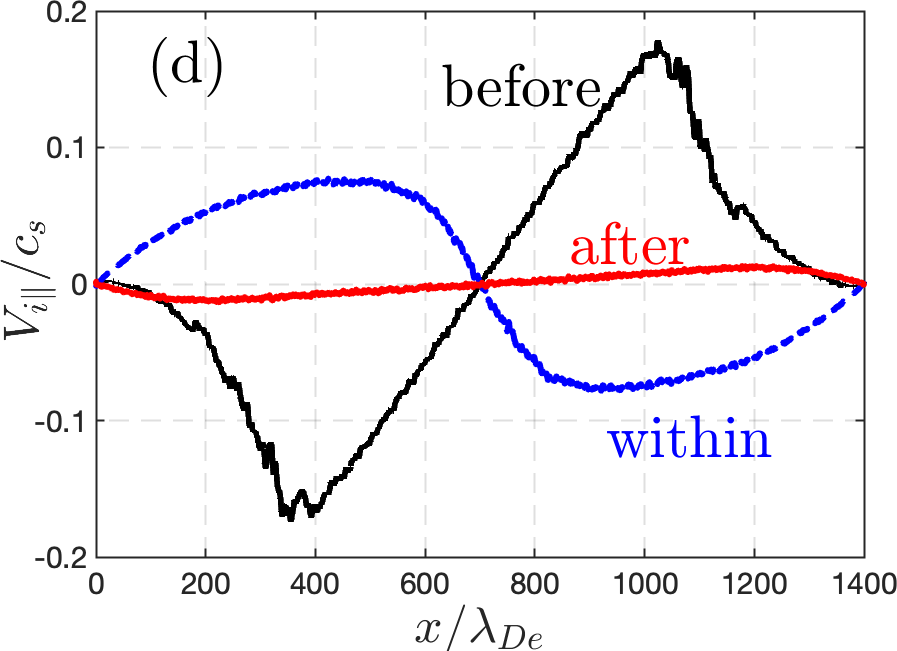}\\
\includegraphics[width=0.4\textwidth]{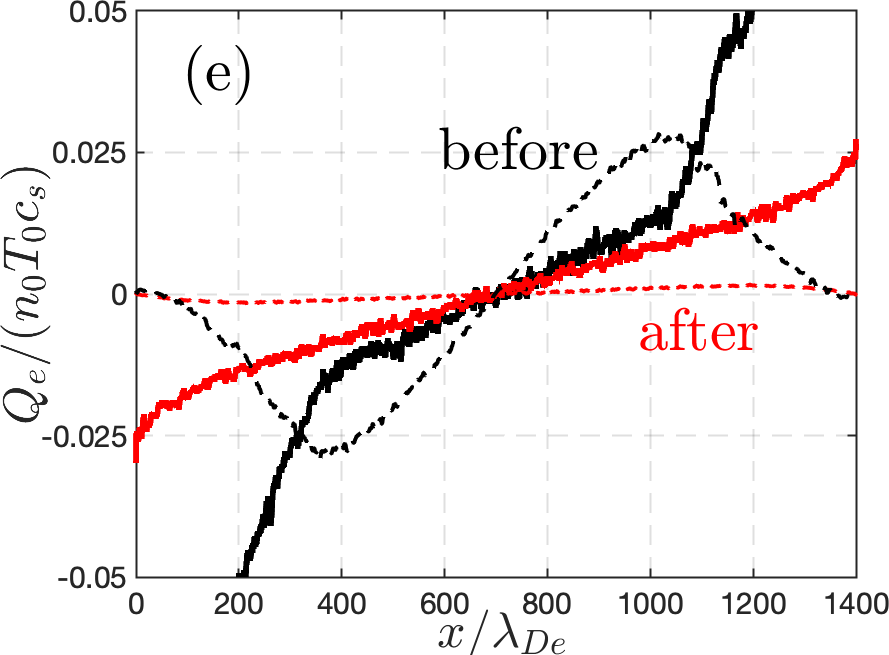}
\caption{Spatial profiles of (a) $n_e$, (b) $T_{e\parallel}$, (c)
  $P=n_eT_{e\parallel}+n_iT_{i\parallel}$ and (d) $V_{i\parallel}$
  before (inside the cooling flow phase), within, and after the
  transition phase from the simulation in
  Fig.~\ref{fig:Te_Ti_para_perp}.  (e) shows the profile of electron thermal conduction (solid thicker lines) and convection
 ($3n_eV_{i\parallel} T_{e\parallel}$, dash thinner lines) fluxes ($Q_e$) before and after the transition phase. Before and after the transition, the thermal conduction is $q_{en}$ and $q_{e\parallel}$ defined in Eq.~(\ref{eq:qn-convective-scaling}) and Eq.~(\ref{eq-qen-Braginskii}), respectively.
\label{fig:Te-ne-Vi-profile}}
\end{figure*}

An even more interesting and impactful result is on $q_{en}$ of a plasma
with thermobath boundary conditions. The subtlety is that cold
electrons from the boundary plasmas can follow the ambipolar electric field into the core plasma. This cold electron beam component was
previously found to restore the free-steaming flux-limiting form of
Eq.~(\ref{eq:qe-free-streaming}) in a semi-infinite plasma~\cite{Zhangfronts}. With cold boundary plasmas on both ends at
$x=\pm L_B,$ the cold electron beams from both sides, accelerated by
the ambipolar electric field, nearly cancel each other, in the
electron flow and thermal conduction flux.  As the result, $q_{en}$
takes on the {\em qualitatively different result by retaining the
  convective energy transport scaling} in the absorbing wall case.
Quantitatively, once the ion recession front, which propagates from
the boundary to the plasma center at the local ion sound speed,
reaches the center, $q_{en}$ has settled into a magnitude that
is solidly {\em sub-dominant} to the actual convective energy flux $3n_e
V_{i\parallel} T_{e\parallel},$ as shown by the black lines in Fig.~\ref{fig:Te-ne-Vi-profile}(e).

Remarkably the core plasma cooling history in the cooling flow phase
is similar if not nearly identical between the absorbing and
thermobath boundary cases.  This is connected to thermal conduction
being sub-dominant to convective energy transport in both cases.  Next, we use the
absorbing wall case to illustrate the characteristics of the cooling
history.  An important result from the VPIC simulations is that the
cooling flow $V_{i\parallel}$ linearly grows in $x$ from zero at $x=0$
to the Bohm speed $u_{Bohm}$ near $x=L_B,$ and it decays in time due
to plasma cooling. This can be understood via a separable solution
$V_{i\parallel}(x,t)=P(x)Q(t)$ to the ion momentum equation, where the
decay of the plasma flow is mostly balanced by the inertial term (e.g., see Fig.~\ref{fig:Te-ne-Vi-profile}),
\begin{align}
\frac{\partial }{\partial t} V_{i\parallel}+
V_{i\parallel}\frac{\partial}{\partial x} V_{i\parallel} = - \nabla p
\approx 0.
\label{eq-ion-momentum2}
\end{align}
The separable solution has
\begin{align}
\frac{dP}{dx}=-\frac{dQ/dt}{Q^2}=C_0,
\end{align}
with $C_0$ a constant.  The solution $P(x) = C_0 x$ underlies
the aforementioned VPIC simulation result.  In the normal situation
with an upstream Maxwellian plasma, $u_{Bohm} \approx
\sqrt{(ZT_{e\parallel} + 3 T_{i\parallel})/m_i}$ because of the
parallel heat flux into the
sheath~\cite{tang2016critical,Yuzhi}. While for the nearly
collisionless plasma in the cooling flow phase during a thermal
quench, the thermal conduction is insignificant, so $u_{Bohm} \approx
\sqrt{(3ZT_{e\parallel} + 3 T_{i\parallel})/m_i}$. Since the cooling
starts at $T_0,$ the Bohm speed at $t=0$ is simply the parallel sound
speed $c_{s} \equiv \sqrt{3(Z+1)T_0/m_i}$ in a magnetized plasma with
strong temperature anisotropy, so we can write $C_0=c_s/L_B$, which
introduces an ion transit time $\tau_{tr}^i=L_B/c_{s}$ for $Q,$
\begin{align}
Q = \left(\delta + \frac{t}{\tau_{tr}^i}\right)^{-1}.\label{eq-Vi-para-scaling}
\end{align}
Here $\delta\sim 1$ depends on the condition when $T_{e\parallel}$
becomes uniform and we assumed $T_{i\parallel}\approx T_{e\parallel}$
due to the convective heat flux dominating the cooling.  Similarly a
separable solution $T_{e\parallel} = M(x) N(t)$ can be found from the
electron energy equation,~\cite{guo-tang-pop-2012a}
\begin{align}
n_e \frac{\partial}{\partial t} T_{e\parallel} +n_e V_{e\parallel}
\frac{\partial}{\partial x} T_{e\parallel}+
2n_eT_{e\parallel}\frac{\partial}{\partial x}V_{e\parallel}
+\frac{\partial}{\partial x} q_{e\parallel} = 0.
  \label{eq-electron-Teparallel}
\end{align}
Ignoring the sub-dominant $\partial q_{en}/\partial x$ term and the
$n_e V_{e\parallel}\partial T_{e\parallel}/\partial x$ term because of
weak $T_{e\parallel}$ variation, one finds
$  {d\ln N(t)}/{dt} = - 2C_0 Q(t) $
for
\begin{align}
\frac{T_{e\parallel}(t)}{T_0}= N(t)=\left(\delta +
\frac{t}{\tau_{tr}^i}\right)^{-2},\label{eq-Te-scaling-convective}
\end{align}
with $M(x)\approx T_0$. This reveals two pieces of interesting and
important physics: (1) the characterized time of collisionless core
$T_{e\parallel}$ cooling is given by the ion sound transit time
$\Delta t_2\sim \tau_{tr}^i$; and (2) $T_{e\parallel}(t)\propto
t^{-2}$ for $t/\tau_{tr}^i\gg 1$ as shown in
Fig.~\ref{fig:Te_Ti_para_perp}. Substitution of
Eq.~(\ref{eq-Te-scaling-convective}) into $u_{Bohm}$ gives
$u_{Bohm}=P(x=L_B)Q$ as expected so that the solutions are
self-consistent.

The temperature $T_2$ signifies the beginning of a transition from
collisionless to collisional cooling, and it is set by the constraint
that the convection flux equals the Braginskii conduction, which
suggests a Knudsen number at the transition
\begin{equation}
    K_{n,2}\sim c_s/v_{th,e}\sim \sqrt{m_e(1+Z)/m_i}.\label{eq-Kn-transition}
\end{equation}
If we ignore the small density variation at the core (see
Fig.~\ref{fig:Te-ne-Vi-profile}) so that $K_{n,2}\approx
K_{n,0}(T_2/T_0)^2$, we find
\begin{equation}
    T_2\sim T_0K_{n,0}^{-1/2}\left[m_e(1+Z)/m_i\right]^{1/4}.\label{eq-T2}
\end{equation}
This agrees well with VPIC simulation data in
Fig.~\ref{fig:Te_Ti_para_perp}, where $T_2\approx 0.06T_0$ from
Eq.~(\ref{eq-T2}) and the simulation. The cooling time from $T_1$ to
$T_2$ is then found from Eq.~(\ref{eq-Te-scaling-convective}) to be
\begin{equation}
    \Delta t_2\sim K_{n,0}^{1/4}\left[m_e(1+Z)/m_i\right]^{-1/8}\tau_{tr}^i.\label{eq-Deltat2}
\end{equation}
For plasma parameters in Fig.~\ref{fig:Te_Ti_para_perp}, $\Delta
t_2\sim 4 \tau_{tr}^i,$ in good agreement
with the simulation data.

{\em Transition phase ($T_{e\parallel} \sim T_2 \sim T_3$ and $\Delta
  t_3 \sim \tau_{tr}^i$):} The cooling flow meets its eventual
collapse when the boundary plasma becomes over-pressured because of
density build-up in the presence of decreasing temperature
gradient. This sets off a recompression of the core that transiently
heats the plasma temperature, especially that of the ions.  Due to the
strong Landau damping of ion-acoustic waves, the recompression flow is
removed over one thermal ion transit period, so $\Delta t_3 \approx
\tau_{tr}^i(T_2)\sim \Delta t_2,$ leaving a core plasma temperature $T_3\approx T_2$ at
the end of the one-bounce transition period, as shown in
Fig.~\ref{fig:Te_Ti_para_perp}.

{\em Collisional cooling phase ($T_{e\parallel}$ from $T_3$ to $T_w,$
  and $\Delta t_4 \gg \tau_{tr}^i$):} Collisional cooling is dominated by the 
thermal conduction $q_{e\parallel}$ given by Braginskii in
Eq.~(\ref{eq-qen-Braginskii}):
\begin{equation}
 3n_e \frac{\partial}{\partial t} T_{e\parallel}
 +\frac{\partial}{\partial x} q_{e\parallel} = 0.
  \label{eq-energy-Braginskii}
\end{equation}
A separable solution $T_{e\parallel}=M(x)N(t)$
with $M(x)\neq const.$ has the core $T_{e\parallel}$ evolve in time as
\begin{equation}
\frac{T_{e\parallel}}{T_3}=\left(\delta_2+1.6K_{n,3}\sqrt{\frac{T_3}{m_e}}\frac{t}
     {L_B}\right)^{-2/5},\label{eq-Tepara-scaling-Braginskii-T3}
\end{equation}
where $\delta_2\sim 1$ accounts for
$\Delta t_2$ and $\Delta t_3$ in $t$. Here all the variables are
normalized using the quantities at the beginning of collisional TQ
including the Knudsen number $K_{n,3}$. Therefore, this
solution can be applied to plasma that is initially within the
collisional regime. For an initially collisionless fusion plasma, we can substitute $T_3\approx T_2$ and hence
$K_{n,3}\approx K_{n,2}$ into
Eq.~(\ref{eq-Tepara-scaling-Braginskii-T3}) to obtain
\begin{equation}
\frac{T_{e\parallel}}{T_3}=\left\{\delta_2+1.6K_{n,0}^{-1/4}\left[\frac{m_e(1+Z)}{m_i}\right]^{1/8}
\frac{t}{\tau_{tr}^i}\right\}^{-2/5}.\label{eq-Tepara-scaling-Braginskii}
\end{equation}
Here $t$ is naturally normalized by $\tau_{tr}^i$ as in
Eq.~(\ref{eq-Te-scaling-convective}) for the cooling flow phase.  The
factor in front has weak powers ($-1/4$ and $1/8$ respectively) of
$K_{n,0}$ and $m_e(1+Z)/m_i,$ so it is also a number of order unity,
like that in Eq.~(\ref{eq-Te-scaling-convective}).  What is different
is that $T_{e\parallel}(t)\propto t^{-2/5}$ as shown in
Fig.~\ref{fig:Te_Ti_para_perp}, in sharp contrast to $T_{e\parallel}(t)\propto t^{-2}$ in the cooling flow phase. This explains a much slower core TQ in the collisional regime than
that in the collisionless regime. Specifically, the core TQ from
$T_3$ to $T_w$ takes
\begin{equation}
    \Delta t_4\sim
    \left(\frac{T_w}{T_0}\right)^{-5/2}K_{n,0}^{-1}\left[\frac{m_e(1+Z)}{m_i}\right]^{1/2}\tau_{tr}^i.\label{eq-Deltat4}
\end{equation}
It is easy to check that $\Delta t_4\gg \Delta t_2$ for $T_w\ll T_2$.

\section{Conclusion and Discussion\label{sec:conclusion}}

In conclusion, the main core thermal collapse ($T_0 \rightarrow T_2
\ll T_0$) is mostly accomplished in the cooling flow phase with the
duration that is a few times the ion sound transit time $\Delta t_2
\sim L_B/c_s,$ due to the surprising physics that electron thermal
conduction is sub-dominant to convective energy transport in a nearly
collisionless cooling plasma.  This is followed by a transition phase
in which over-pressured boundary plasma reheats the core plasma by
compression. The final collisional cooling phase takes so long that
the experimentally observed deep cooling to tens or a few eVs in
milliseconds or shorter time must be due to a different mechanism,
with impurity radiation a leading candidate. The source of these
impurities, in naturally occurring disruptions, is likely the result
of intense plasma-material interaction in the cooling phases preceding
the final collisional stage, during which the outgoing plasma power flux is
far greater.

\begin{figure}[h!]
\centering
\includegraphics[width=0.45\textwidth]{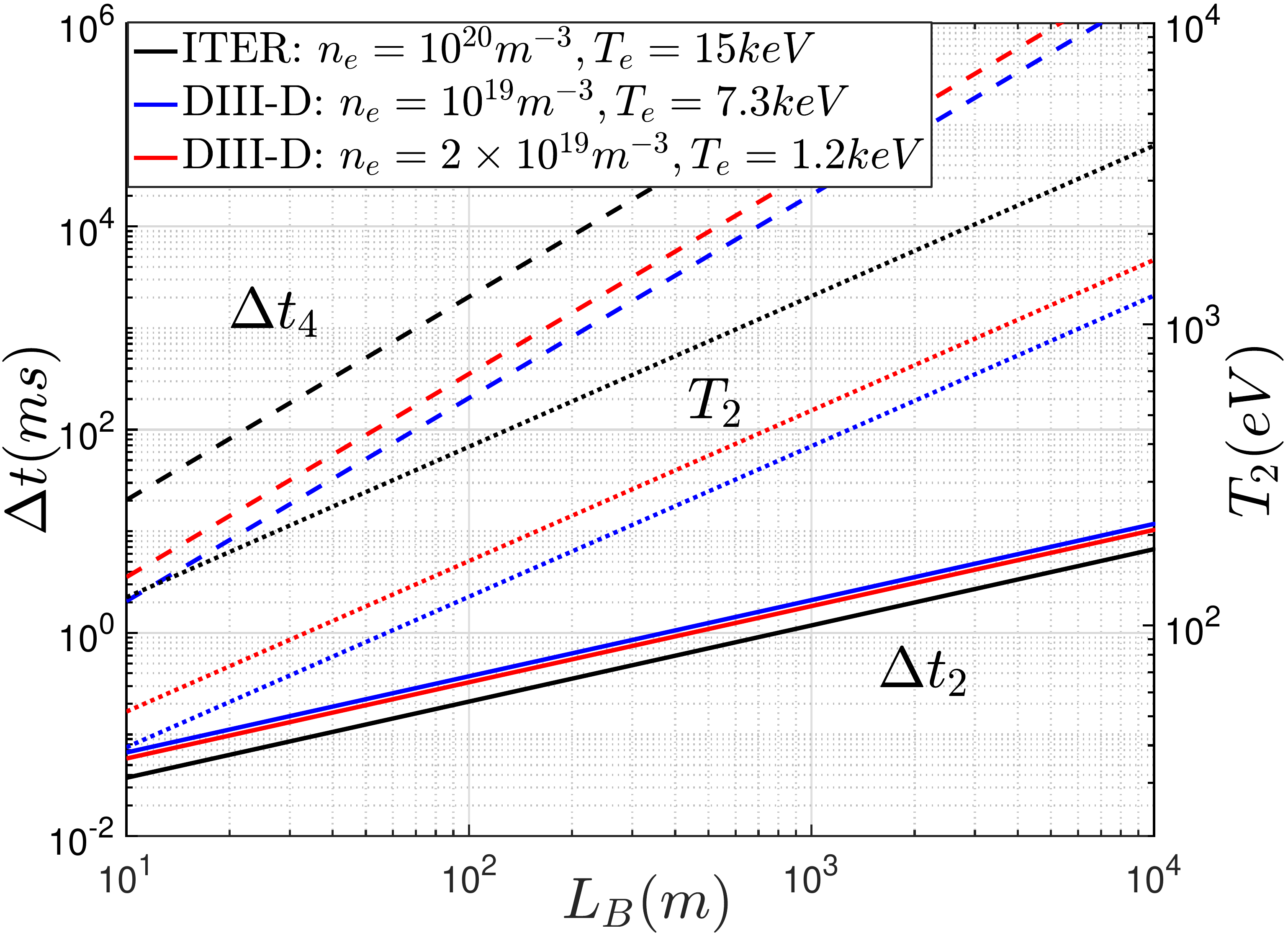}
\caption{$\Delta t_{2,4}$ from
  Eqs.~(\ref{eq-Deltat2},\ref{eq-Deltat4}) (solid and dash lines for
  left y-axis) and $T_2$ from Eq.~(\ref{eq-T2}) (dotted lines for
  right y-axis) of representative DIII-D and ITER plasmas. Here we
  choose hydrogen ions and $T_w=10eV$.}
\label{fig:TQ-durations}
\end{figure}

The physics scaling reveals that for a fusion-grade plasma in which
$T_0$ and $n_0$ are severely constrained, the variability in TQ
duration, which experimentally is associated with $\Delta t_2,$ is
mostly set by the magnetic connection length $L_B.$ This potentially
provides a means to determine the otherwise difficult-to-access $L_B$
of globally stochastic magnetic fields. To quantify our findings, we
show $\Delta t_{2,4}$ from Eqs.~(\ref{eq-Deltat2},\ref{eq-Deltat4})
and $T_2$ from Eq.~(\ref{eq-T2}), of representative DIII-D and ITER
plasmas for a range of $L_B$ in Fig.~\ref{fig:TQ-durations}. This
provides a baseline prediction of TQ duration for comparison with
experimental measurements and more involved transport calculations.
For all cases considered, it is striking that the duration of the
collisional cooling phase, $\Delta t_4,$ is orders of magnitude longer
than that of the collisionless cooling phase $\Delta t_2.$
Furthermore, for longer magnetic connection length $L_B,$ the ratio of
$\Delta t_4$ and $\Delta t_2$ becomes larger. The transition
temperature $T_2$ that separates the collisionless and collisional
cooling phases becomes lower if a stronger field line stochasticity
produces a shorter magnetic connection length $L_B.$

Since the analysis given in this paper focuses on the regime of
parallel transport dominating over perpendicular transport, the
experimentally applicable regime for the predicted results of $\Delta
t_{2,4}$ and $T_2$ is toward the end of short magnetic connection
length $L_B,$ for which $\Delta t_{2,4}$ and $T_2$ are of smaller
values. For $L_B \sim 10^3$~m, parallel transport can certainly
produce a core temperature collapse on the order of a millisecond or
so.  The main thermal quench time ($\Delta t_2$) becomes shorter if
the magnetic connection length is further reduced, reaching $\Delta
t_2 < 0.1$~ms for $L_B\sim 10$~m. Such a short magnetic connection
length can be experimentally realized by the injected high-Z pellets that
directly intersect the core plasma, even with the nested flux surfaces
remaining intact.

\acknowledgments

We thank the U.S. Department of Energy Office of Fusion Energy
Sciences and Office of Advanced Scientific Computing Research for
support under the Tokamak Disruption Simulation (TDS) Scientific
Discovery through Advanced Computing (SciDAC) project at Los Alamos
National Laboratory (LANL) under contract
No. 89233218CNA000001. Y.Z. was supported under a Director’s
Postdoctoral Fellowship at LANL. This research used resources of the
National Energy Research Scientific Computing Center (NERSC), a
U.S. Department of Energy Office of Science User Facility operated
under Contract No. DE-AC02-05CH11231, and the Los Alamos National
Laboratory Institutional Computing Program, which is supported by the
U.S. Department of Energy National Nuclear Security Administration
under Contract No. 89233218CNA000001.

\appendix

\section{Derivation of collisional cooling scaling in Eq.~(\ref{eq-Tepara-scaling-Braginskii-T3})\label{appendix:collisional-cooling}}

In a strong collisional plasma, only considering the thermal flux in the energy equation provides~\cite{braginskii}
\begin{align}
\label{eqn:start_Te}
3n_e\frac{\partial T_e}{\partial t} = - \frac{\partial q_e}{\partial x} =  \frac{\partial }{\partial x}(\kappa_{||}\frac{\partial T_e}{\partial x}), 
\end{align}
where
\begin{align}
\kappa=3.16\frac{n_eT_e\tau_{e}}{m_e},
\end{align}
and we ignored the subscript $\parallel$ in $T_e$ and $\kappa$. Note that we have summed up the energy equations of electrons and ions and kept only the electron thermal flux. It is also worth noting that $n_e\tau_{e}/T_e^{3/2}$ is a constant when ignoring the dependence of $\ln \Lambda$ on the density and temperature so that we obtain 
\begin{align}
\frac{\partial T_e}{\partial t}&= \frac{3.16}{3}\frac{2}{7}\frac{\tau_{e}}{m_eT_e^{3/2}} \frac{\partial^2 }{\partial x^2}T_e^{7/2}.
\end{align}
Now if we ignore the density variation in space and time, we arrive at the heat equation with a constant coefficient $C_0 \approx 0.3 \lambda_{emfp,3}v_{th,3}/T_3^{5/2}$, where $v_{th,3}=\sqrt{T_3/m_e}$ and $\lambda_{emfp,3}$ are the electron thermal speed and mean free path at the temperature $T_3$ (note that $T_3$ is used for the normalization of the electron temperature).

A separable solution $T_e = M(x)N(t)$ would yield
\begin{align}
\frac{dN/dt}{C_0N^{7/2}}=\frac{1}{M} \frac{\partial^2 M^{7/2}}{\partial x^2} = const. \equiv C_1 <0, 
\end{align}
from which we obtain
\begin{equation}
N=\left(N_0^{-5/2}-2.5C_0C_1t\right)^{-2/5},\label{eqn-N}
\end{equation}
where $N_0 = N(t=0)$. For the equation of $M(x)$, we define $\Upsilon=M^{7/2}$, which satisfies
\begin{align}
\label{eqn:dPdx2}
\frac{\partial^2 \Upsilon}{\partial x^2} = C_1\Upsilon^{2/7}.
\end{align}
Such an equation has a solution of the implicit form
\begin{align}
\label{eqn:HP}
\frac{H(C_1,C_2,\Upsilon)^2\Upsilon^2}{C_2}=(x+C_3)^2
\end{align}
where $C_2$ and $C_3$ are constants, and 
\begin{align}
H(C_1,C_2,\Upsilon)=_{2}F_1\left(\frac{1}{2}, \frac{7}{9}, \frac{16}{9}, -\frac{14C_1\Upsilon^{9/7}}{9C_2}\right).
\end{align}
is a hypergeometric function~\cite{Brychkov2008}.

Now we consider the boundary conditions to settle down $C_{1,2,3}$, where we let $N$ be dimensionless such that $M(x=0)=T_3$ at the center of the simulation domain. As a result, $\Upsilon (x=0)= M(x=0)^{7/2}=T_3^{7/2}$. We assume that the wall temperature is negligibly small $T_w\approx 0$ so that $\Upsilon (x=L_B) =0$. These conditions turn into 
\begin{align}
C_3=-L_B, \\
H\left(C_1,C_2,T_3^{7/2}\right)^2T_3^7=C_2L_B^2.
\label{eqn:HC3}
\end{align}
To solve $C_1$ and $C_2$ from Eq.~(\ref{eqn:HC3}), we need more conditions, which are $\frac{dM}{dx}|_{x\rightarrow 0} \rightarrow 0$ due to the system symmetry and $\frac{dM}{dx} <0$ at $x>0$. Recalling $C_1<0$, these conditions together with Eq.~(\ref{eqn:HC3}) provide $C_1\rightarrow -2.13 T_3^{5/2}/L_B^2$ and $C_2=3.31T_3^7/L_B^2$. As a result, Eq.~(\ref{eqn-N}) becomes
\begin{equation}
\frac{T_{e}}{T_3}=\left(N_0^{-5/2}+1.6v_{th,3}\frac{\lambda_{emfp,3}}{L_B}\frac{t}
     {L_B}\right)^{-2/5}
     \label{eq-Tepara-scaling-Braginskii-T3_methods}
\end{equation}
which is Eq.~(\ref{eq-Tepara-scaling-Braginskii-T3}) in the main text.

\bibliography{main}

\end{document}